\documentclass[fleqn,12pt,twoside]{article}
\usepackage{espcrc1}

\usepackage{graphicx}
\usepackage[figuresright]{rotating}

\newcommand{\Bx}{x_{\rm B}}

\newcommand{\AmS}{{\protect\the\textfont2
  A\kern-.1667em\lower.5ex\hbox{M}\kern-.125emS}}

\title{Extraction of the Generalized Parton Distribution  $H(\xi,\xi,t)$ 
from DVCS}

\author{\underline{V.A. Korotkov}\address{IHEP, Protvino, RU-142281, Russia},
        W.-D. Nowak\address{DESY, Zeuthen, D-15738, Germany}}

\begin{document}

% typeset front matter
\maketitle

\begin{abstract}
A simple way to extract the Generalized Parton Distribution 
$H(\xi,\xi,t)$ at small $t$ from the single beam-spin asymmetry in
Deeply Virtual Compton Scattering is proposed. Projections are given for
future measurements at the HERMES experiment upgraded with a recoil
detector.
\end{abstract}

\section{Introduction}
First results on measurements of the single beam-spin asymmetry in 
Deeply Virtual Compton Scattering (DVCS) have recently been published 
by the HERMES \cite{dvcshermes} and CLAS \cite{dvcsclas} collaborations.
Both collaborations will upgrade their apparatus to measure the 
DVCS process with improved accuracy. 

There are four different types 
of twist-2 quark Generalized Parton Distributions (GPDs) contributing
to the DVCS process. In the unpolarized distributions,
$H^q(x,\xi,t)$ and $E^q(x,\xi,t)$, the quark helicities are
summed over. The polarized distributions, $\widetilde{H}^q(x,\xi,t)$ 
and $\widetilde{E}^q(x,\xi,t)$, are responsible for the differences 
between right- and left-handed quarks. The GPDs depend on two 
longitudinal momentum fraction variables $x$ and $\xi$ (skewedness) 
and the squared momentum transfer $t$ between initial and final nucleon 
states. The cross-section of the DVCS process and its interference with the 
Bethe-Heitler (BH) process has been considered in a number of papers 
\cite{Ji,rad,diehl,belitsky}. The detailed form of the Compton amplitude 
can be found in Ref.~\cite{jbl2}.
The process amplitude ${\cal T}$ is the sum of the DVCS 
and BH amplitudes, ${\cal T}_{\rm DVCS}$ and ${\cal T}_{\rm BH}$. Their 
interference opens the possibility to access the DVCS amplitudes 
\cite{diehl} through a measurement of the interference term 
${\cal I} = {\cal T}_{\rm DVCS} {\cal T}_{\rm BH}^\ast
+ {\cal T}_{\rm DVCS}^\ast {\cal T}_{\rm BH}$.
Explicit expressions for the amplitudes of the DVCS, Bethe-Heitler and
interference terms including the first subleading correction in $1/Q$
have been calculated in Ref.~\cite{belitsky} for different kinds
of polarized and unpolarized initial particles.
The four-fold cross-section for the process 
$e (k) p (P_1) \to e^\prime (k^\prime) p^\prime (P_2) \gamma (q_2)$ with
an unpolarized proton target depends on the Bjorken variable $\Bx$, 
the lepton energy fraction $y = P_1\cdot q_1/P_1\cdot k$, 
$t = (P_2 - P_1)^2$, and the
azimuthal angle $\phi$ between the lepton and hadron scattering planes:
\begin{eqnarray}
\frac{d\sigma}{d\Bx dy d|t| d\phi}
=
\frac{\alpha^3  \Bx y } {8 \, \pi \,  Q^2 \sqrt{1 + 4 \Bx^2 M^2/Q^2}}
\left| \frac{\cal T}{e^3} \right|^2 \, .
\label{eq:xsection}
\end{eqnarray}
Here, $Q^2 = - q_1^2$ and $q_1 = k - k'$.
The squared amplitudes $|{\cal T}_{\rm BH}|^2$ and $|{\cal T}_{\rm DVCS}|^2$,
and the interference term ${\cal I}$ are expressed in terms of finite sums 
of Fourier harmonics with respect to the azimuthal angle $\phi$ (see Eq.s
(25-27) in Ref.~\cite{belitsky}). Note that for the DVCS process the 
skewedness variable can be determined by $\xi = \Bx/(2 - \Bx)$.

\section{Projection for a future DVCS measurement at HERMES}
The HERMES collaboration will upgrade their apparatus with a recoil detector
\cite{prchermes} to measure the single beam-spin and the beam-charge 
asymmetries using an unpolarized proton target. Prospects for a measurement 
of the single beam-spin asymmetry are investigated in this paper. For the 
projection of the expected statistical accuracy the theoretical approach of
Ref.~\cite{belitsky} was adopted. In the analysis described below five 
different versions of parameterizations for twist-2 GPDs
are used. They are referred to as (A) to (E) (for details see 
Ref.~\cite{dvcsepjc}). The twist-3 GPDs were taken to be zero for simplicity.

The projections of the statistical accuracy for measurements of 
the single beam-spin asymmetry at HERMES were calculated for an integrated 
luminosity of $2$~fb$^{-1}$, corresponding to one nominal year of data taking.
The HERMES acceptance for the detection
of the scattered electron, the real gamma and the recoil proton 
has been taken into account. The following kinematic cuts were applied:
$E_{e^\prime} > 3.5$~GeV, $E_\gamma > 0.8$~GeV, $P_{p'} > 0.2$~GeV/c, 
$W^2 > 4$~GeV$^2$, $Q^2 > 1$~GeV$^2$, and 
$15 < \Theta_{\gamma \gamma^*} < 70$~mrad. Here, $W^2 = (P_2 + q_2)^2$ and
$\Theta_{\gamma \gamma^*}$ is the polar angle between virtual and real photons.

 The remarkably improved accuracy in measuring the variable $t$ with
the recoil detector will allow accurate studies of kinematic dependences
of the asymmetry. For this purpose it is convenient to define
an appropriate $\sin \phi$ moment\footnote{Note that a
factor $2$ is used in Eq.~(\ref{eq:sinmomdef}) to obey
to the HERMES definition of the moments \cite{dvcshermes}}
of the asymmetry:
\begin{equation}
A_{LU}^{\sin \phi} =  2 \cdot {{\int_0^{2 \pi} d\phi \, \sin \phi 
          \{ d \sigma (\overrightarrow{e} p) / d\phi \, - \,
             d \sigma (\overleftarrow{e} p) / d\phi \} }        \over
          { \int_0^{2 \pi} d\phi \, 
            \{ d \sigma (\overrightarrow{e} p) / d\phi \, + \,
             d \sigma (\overleftarrow{e} p) / d\phi \} }}
\label{eq:sinmomdef}
\end{equation}

\begin{figure}[hbt]\centering
\vspace*{-0.5cm}
\includegraphics[width=7.9cm]{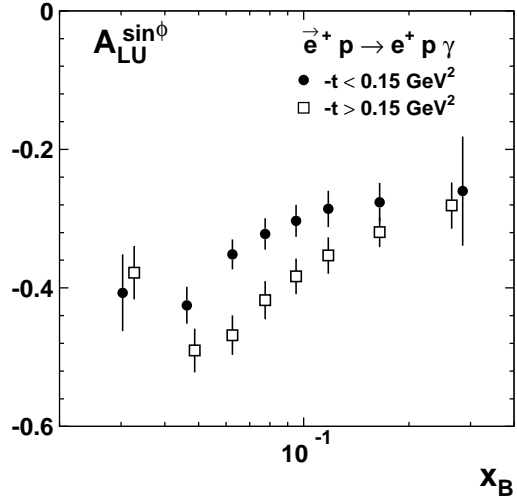}
\vspace*{-1.5cm}
\caption{\it Projected statistical accuracy for a future HERMES measurement 
of the $A_{LU}^{\sin \phi}$ moment of the DVCS single beam-spin asymmetry 
as a function of $\Bx$ for two regoins in $-t$. The calculations,
using GPD model version (E) (see Ref.~\cite{dvcsepjc}), are based on an 
integrated luminosity of 2 fb$^{-1}$.}
\label{fig:momxdep}
\end{figure}

The projections for the statistical accuracy of the moment
$A_{LU}^{\sin \phi}$ as a function of $\Bx$ for two
regions in $t$ are shown in Fig.~\ref{fig:momxdep}.

\section{Extraction of the Generalized Parton Distribution  $H(\xi,\xi,t)$}

The asymmetry moment depends on all four DVCS amplitudes, $\cal{H}, 
\widetilde {\cal H}, \cal{E}, \widetilde {\cal E}$, which embed the four
Generalized Parton Distributions $H^q(x,\xi,t), E^q(x,\xi,t), 
\widetilde{H}^q(x,\xi,t), \widetilde{E}^q(x,\xi,t)$ (see e.g. 
Ref.~\cite{dvcsepjc}). A general method for a simultaneous extraction of 
all four GPDs from the asymmetry moments has 
not been worked out, yet. As will be shown in the following, information 
on the imaginary part of the DVCS amplitude $\cal H$, ${\rm Im } \, {\cal H}$, 
can nevertheless be obtained under quite reasonable assumptions from future 
measurements of the $A_{LU}^{\sin \phi}$ moment at HERMES, as outlined above.
The main part of the DVCS cross section contributing to 
the denominator of Eq.~(\ref{eq:sinmomdef}) is proportional to
\begin{eqnarray}
{\cal C}^{\rm DVCS}_{\rm{unp}}
\!\!\!&=&\!\!\!
\frac{1}{(2 - \Bx)^2}
\Bigg\{
4 (1 - \Bx)
\left(
{\cal H} {\cal H}^\ast
+
\widetilde{\cal H} \widetilde {\cal H}^\ast
\right)- \Bx^2
\bigg(
{\cal H} {\cal E}^\ast
+ {\cal E} {\cal H}^\ast
+ \widetilde{{\cal H}} \widetilde{{\cal E}}^\ast
+ \widetilde{{\cal E}} \widetilde{{\cal H}}^\ast
\bigg)
\nonumber\\
&&\qquad\qquad\;
-
\left( \Bx^2 + (2 - \Bx)^2 \frac{t}{4M^2} \right)
{\cal E} {\cal E}^\ast
- \Bx^2 \frac{t}{4M^2}
\widetilde{{\cal E}} \widetilde{{\cal E}}^\ast
\Bigg\}.
\label{eq:denominator}
\end{eqnarray}
The numerator of Eq.~(\ref{eq:sinmomdef}) is proportional to 
${\rm Im} \ {\cal C}^{\cal I}_{\rm unp}$, where
\begin{equation}
 {\cal C}^{\cal I}_{\rm unp}
= F_1 {\cal H} + \frac{\Bx}{2 - \Bx}(F_1 + F_2) \widetilde {\cal H} - 
\frac{t}{4M^2} F_2 {\cal E}
\label{eq:numerator}
\end{equation}
Here, $F_1$ and $F_2$ are the Dirac and Pauli form factors. From the
expressions (\ref{eq:denominator}) and (\ref{eq:numerator}) it may be seen
that at small values of $t$ and $\Bx$ the main contributions  come from 
the first terms of both expressions. This observation can be checked 
numerically using different models for GPDs. In the following the region 
of small $-t$ ($<$ 0.15~GeV$^2$) is considered.

\begin{figure}[h]\centering
\vspace*{-1.0cm}
\includegraphics[width=14cm]{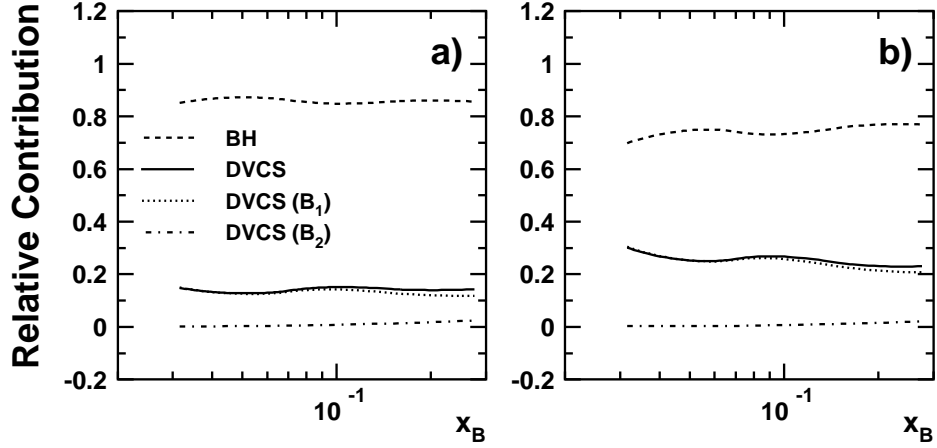}
\vspace*{-1.5cm}
\caption{\it Relative contribution of the BH and DVCS cross sections to the 
denominator of Eq.~(\ref{eq:sinmomdef}). The relative contributions coming from 
the DVCS cross section parts proportional to $B_1$ and $B_2$ are shown also. Results 
for the GPD versions (A) and (B) are shown in panels a) and b), respectively. }
\label{fig:contrsig}
\end{figure}

\begin{figure}[h]\centering
\vspace*{-1.0cm}
\includegraphics[width=14cm]{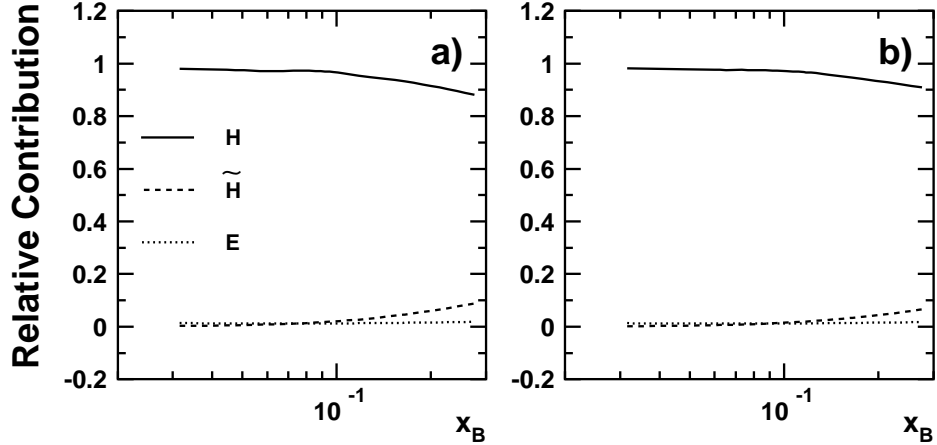}
\vspace*{-1.5cm}
\caption{\it Relative contribution of terms from Eq.~(\ref{eq:numerator}) induced 
to the numerator of Eq.~(\ref{eq:sinmomdef}) due to the amplitudes $\cal H$,
$\widetilde{\cal H}$, and $\cal E$. Results for GPD versions (A) and (B) are shown in
panels a) and b), respectively.}
\label{fig:contrint}
\end{figure}

Expression (\ref{eq:denominator}) can be regrouped into two terms,
${\cal C}^{\rm DVCS}_{\rm{unp}} \, = \, B_1 \, + \, B_2$, with
\begin{equation}
B_1 \, = \, 4 \, \frac{ 1 - \Bx }{(2 - \Bx)^2} \bigg( {\rm Im} {\cal H} \bigg)^2
\end{equation}
and $B_2$ representing the remainder of ${\cal C}^{\rm DVCS}_{\rm{unp}}$.
The relative contributions of the BH and DVCS cross sections to the denominator of 
Eq.~(\ref{eq:sinmomdef}) are shown in Fig.~\ref{fig:contrsig}. The same figure shows
the relative contributions coming from the DVCS cross section parts proportional to
$B_1$ and $B_2$. From the figure it may be concluded that in the kinematics
considered in this paper the main component of the 
DVCS cross section contributing to the denominator of Eq.~(\ref{eq:sinmomdef})
is due to ${\rm Im} {\cal H}$.
In Fig.~\ref{fig:contrint} are shown the relative contributions of the three terms 
from Eq.~(\ref{eq:numerator}), induced to the numerator of Eq.~(\ref{eq:sinmomdef}) 
due to the amplitudes $\cal H$, $\widetilde{\cal H}$ and $\cal E$ . Again, it may 
be concluded that the main contribution comes from ${\rm Im} {\cal H}$.

\begin{figure}[h]\centering
\vspace*{-1.5cm}
\includegraphics[width=14cm]{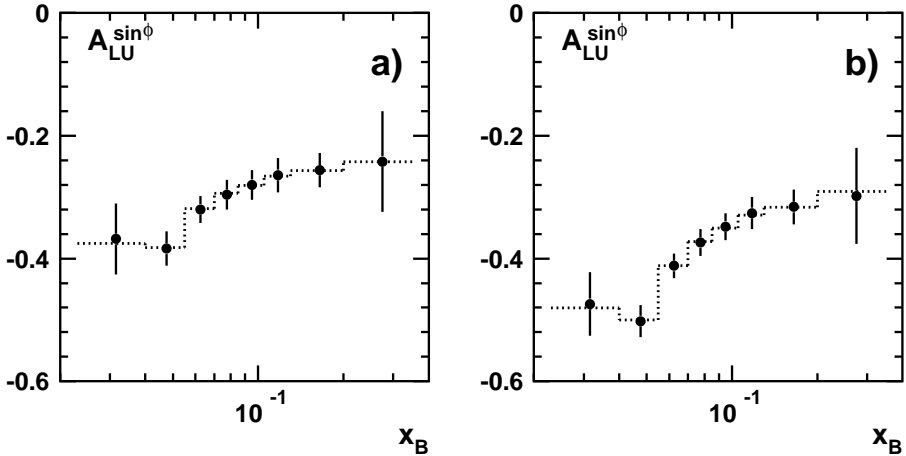}
\vspace*{-1.5cm}
\caption{\it Results of a fit to the projected moments for GPD model versions 
(A) and (B) on the basis of Eq.s~(\ref{eq:momparam}) and (\ref{eq:hparam}),
calculated for $-t < 0.15$ GeV$^2$.}
\label{fig:fits}
\end{figure}

These numerical studies 
show that the most significant contribution to $A_{LU}^{\sin \phi}$ at small $-t$ 
($<$ 0.15~GeV$^2$) is originating from ${\rm Im } \, {\cal H}$. The sum of all 
remaining contributions amounts only to a few percent. This conclusion has been 
checked for all five GPD model versions considered in Ref.~\cite{dvcsepjc}. 
Therefore, the assumption, that the $A_{LU}^{\sin \phi}$  
moment may be analysed as a function of ${\rm Im } \, {\cal H}$ only, appears to be 
quite reasonable.
In this scenario, the moment (\ref{eq:sinmomdef}) can be expressed in the
following form:
\begin{equation}
A_{LU}^{\sin \phi} = 2 \cdot {{{\int d\Omega \, \sin \phi \,
           {d \sigma^I_0}/{d\Omega} \cdot Im {\cal H}
               }}        \over
          { \int d\Omega \,  {d \sigma^{BH}}/{d\Omega} \, + \,
            \int d\Omega \,  {d \sigma^{DVCS}_0}/{d\Omega}  
             \cdot (Im {\cal H})^2  }} \ .
\label{eq:momparam}
\end{equation}
Here, $\Omega$ denotes all relevant variables, ${d \sigma^{BH}}/{d\Omega}$ is the 
cross section of the BH process, $d \sigma^{DVCS}_0 / d\Omega$ is that of
the unpolarized DVCS process, and $d \sigma^I_0 / d\Omega$ is that of the
interference term. Both $d \sigma^I_0 / d\Omega$ and 
$d \sigma^{DVCS}_0 / d\Omega$ are calculated assuming 
$ Im {\cal H} = 1, \, Re {\cal H} = 0, \,
 \widetilde{\cal H} = 0, \, {\cal E} = 0, \, \widetilde{\cal E} = 0$. Based on 
this assumption a procedure has been developed to extract ${\rm Im } \, {\cal H}$ 
in a parameterized form from future HERMES measurements.
A simple parameterization with two free parameters  $c_1$ and $c_2$,
\begin{equation}
 {\rm Im } \, {\cal H} (\xi) = c_1 \xi^{c_2},
\label{eq:hparam}
\end{equation}
was adopted following an observation made in Ref.~\cite{freundmc}.
Fits to the projected moment asymmetry in the region $-t < 0.15$ GeV$^2$,
performed on the basis of Eq.s~(\ref{eq:momparam}) 
and (\ref{eq:hparam}), show good descriptions of the $\Bx$-dependence
of $A_{LU}^{\sin \phi}$. The results of the fits to the projected moments for GPD 
model versions (A) and (B), taken as examples, are shown in Fig.~\ref{fig:fits}. 
Typical values of $\chi^2/ndf$ are $(4 \div 6) /6$ if every point is varied in 
limits of $\pm 3 \sigma$ of its projected error. Deviations of the extracted function
${\rm Im } \, {\cal H}$ from that used as input for the projections are 
found to be small within the expected statistical accuracy. 

\begin{figure}[ht]\centering
\vspace*{-0.5cm}
\includegraphics[width=10.7cm]{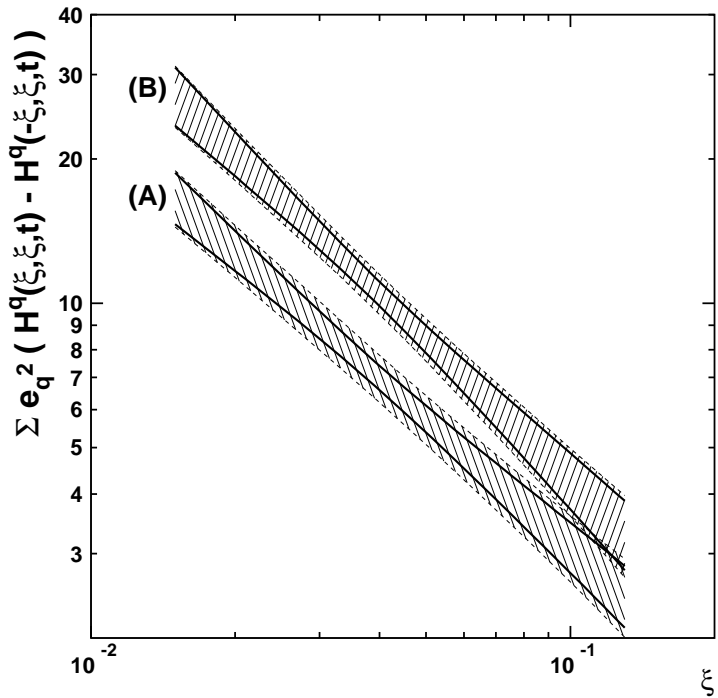}
\vspace*{-0.5cm}
\caption{\it Projection for a future HERMES measurement, based on 2 fb$^{-1}$, 
of $ \sum_q e^2_q \bigl( H^q( \xi, \xi, t) - H^q( -\xi, \xi, t) \bigr)$, shown
in dependence of $\xi$ and $\Bx$ simultaneously.
The fully correlated 1$\sigma$ error bands (solid lines) show the
projected statistical accuracy for the two GPD model versions (A) and (B),
chosen as examples (see Ref.~\cite{dvcsepjc}). The hatched bands represent the
linear sum of the statistical 1$\sigma$ error and a systematic error estimated 
to be due to the 
approximations inherent in the used analysis method.
}
\label{fig:hband}
\end{figure}

To leading order in $\alpha_s$, the determination of ${\rm Im } \, {\cal H}$ 
corresponds to a measurement of the 
quantity $ \sum_q e^2_q \bigl( H^q( \xi, \xi, t) - H^q( -\xi, \xi, t) \bigr)$,
taken along the diagonals $x = \pm \xi$ (cf. e.g. Eq.~(19) in Ref.~\cite{dvcsepjc}). 
The fully correlated 
1$\sigma$ error bands for a future measurement of this quantity at HERMES
in the region $-t < 0.15$ GeV$^2$, as obtained after the fit, are shown in 
Fig.~\ref{fig:hband} for GPD model versions (A) and (B).
The systematic uncertainty of this simple analysis method, as depicted in 
Fig.~\ref{fig:hband}, was estimated to be the average of the deviations
between reconstructed and input functions for all versions of GPDs and 
was found to be smaller than the envisaged statistical uncertainty.

\section{Conclusions}
Numerical studies carried out for different models of GPDs have shown that the most 
significant contribution to $A_{LU}^{\sin \phi}$ at small $-t$ and $0.03 < x_B < 0.3$
is originating from 
${\rm Im } \, {\cal H}$. This observation allows to develop a procedure for the
extraction of the Generalized Parton Distribution $H$ from future HERMES 
measurements of the single beam-spin asymmetry. A simple parameterization 
of ${\rm Im } \, {\cal H}$ with two free parameters allows to describe a dependence of
$A_{LU}^{\sin \phi}$ on $\Bx$ at small $t$. Deviations of the extracted function
${\rm Im } \, {\cal H}$ from that used as input for the projections are 
small inside the expected statistical accuracy.

We are grateful to M.~Diehl and R.~Kaiser for a careful reading of the
manuscript.

\end{document}